\documentclass[12pt,letterpaper,fleqn]{article}
\usepackage{comment,etex,authblk,amsmath,fullpage,amsfonts,titletoc,titlesec,amssymb,theorem}
\usepackage[onehalfspacing]{setspace}
\usepackage{graphicx,hyperref,pgfplots,dcolumn,ctable,floatpag,ulem}
\usepackage{array,multirow,float,subcaption,algorithm, algorithmic}

\usepackage{bm}

\newlength\mytemplength

\titleformat*{\section}{\large\bfseries}
\titleformat*{\subsection}{\normalsize\bfseries}
\usepackage{graphicx}

\hypersetup{colorlinks=true,
linkcolor=blue,
citecolor=blue,
urlcolor=blue}

\usetikzlibrary{pgfplots.groupplots}
\usetikzlibrary{patterns, arrows}
\usetikzlibrary{arrows,positioning,shadows,shapes,calc,fit,backgrounds}
\tikzset{>=stealth}

\reversemarginpar

\floatpagestyle{empty}

\makeatletter
\def\pgfplots@drawtickgridlines@INSTALLCLIP@onorientedsurf#1{}
\makeatother

\usepackage{caption}

\usepackage{natbib}
\makeatletter
\renewcommand{\bibsection}{
\begin{center}
\section*{\refname\@mkboth{\MakeUppercase{\refname}}
{\MakeUppercase{\refname}}}
\end{center}
}
\makeatother

\newtheorem{theorem}{Theorem}

\newtheorem{definition}{Definition}

\newtheorem{theorem-app}{Theorem}[section]
\newtheorem{lemma-app}[theorem-app]{Lemma}
\newtheorem{proposition-app}[theorem-app]{Proposition}
\theorembodyfont{\upshape}

\usepackage{natbib}
\makeatletter
\renewcommand{\bibsection}{
\begin{center}
\section*{\refname\@mkboth{\MakeUppercase{\refname}}
{\MakeUppercase{\refname}}}
\end{center}
}
\makeatother

\newenvironment{proof}[1][\proofname]{
\par\normalfont\trivlist\item[\hskip\labelsep\textbf{#1}.]\ignorespaces}
{\hfill $\square$ 
\endtrivlist}
\newcommand{\proofname}{Proof}

\usepackage{bbm}
\usepackage{color}
\usepackage[dvipsnames]{xcolor}

\usepackage{mathtools}

\newcommand{\be}{\begin{equation}}

\newcommand{\ee}{\end{equation}}
\newcommand{\bea}{\begin{eqnarray}}
\newcommand{\eea}{\end{eqnarray}}
\newcommand{\bee}{\begin{equation*}}
\newcommand{\eee}{\end{equation*}}

\begin{document}
\title{Analysis of Contagion in China's Stock Market: A Hawkes Process Perspective}

\author{Junwei Yang\footnote{Junwei Yang is at school of statistics and data science, Shanghai University of Finance and Economics(SUFE). I would like to thank my supervisor Professor Hongbiao Zhao and members of my dissertation committee, Professor Ning Chang, Professor Jianhua Hu, Professor Xin He for their helpful comments and advice. I also thank Professor Haibo Shi and Professor Yuan Zhang, for their kind guidance and useful discussion. Any errors or omissions are the sole responsibility of the author.}}

\date{\begin{footnotesize}This version: May 18, 2023 \end{footnotesize}}

\maketitle

\begin{abstract} This study explores contagion in the Chinese stock market using Hawkes processes to analyze autocorrelation and cross-correlation in multivariate time series data. We examine whether market indices exhibit trending behavior and whether sector indices influence one another. By fitting self-exciting and inhibitory Hawkes processes to daily returns of indices like the Shanghai Composite, Shenzhen Component, and ChiNext, as well as sector indices (CSI Consumer, Healthcare, and Financial), we identify long-term dependencies and trending patterns, including upward, downward, and oversold rebound trends. Results show that during high trading activity, sector indices tend to sustain their trends, while low activity periods exhibit strong sector rotation. This research models stock price movements using spatiotemporal Hawkes processes, leveraging conditional intensity functions to explain sector rotation, advancing the understanding of financial contagion.

\noindent

\bigskip\bigskip\bigskip

\noindent
\begin{small}Keywords:\ Hawkes Process, Financial Contagion, Sector Rotation
\\

\end{small}

\end{abstract}

\renewcommand{\thefootnote}{\number\value{footnote}}

\pagenumbering{arabic}
\def\baselinestretch{1.617}\small\normalsize%

\clearpage

\section{Introduction}

\subsection{Research Background}
The 2008 subprime crisis triggered a "tsunami" in the U.S. financial markets, leading to a domino effect. Financial contagion focuses on discussing the interactions between different entities under the special structure of financial markets. Contagion manifests in different patterns, including temporal dependencies, such as seasonal river water levels, earthquakes and their aftershocks, and spatial dependencies, such as social network relationships and supply chain relationships. In the stock market, the phenomenon of "sector rotation" describes the alternating rise and fall of different industry stock prices, showing temporal differences and spatial rotation characteristics similar to the temporal and spatial contagion mentioned above. A class of stochastic processes called Hawkes processes provides tools for characterizing the contagion of risk events and has been widely applied in fields such as earthquake prediction. Introducing Hawkes processes into financial markets to describe the interactions of risk, returns, volatility, and other elements between financial markets, institutions, and assets provides a new perspective for explaining financial market. This paper uses multivariate Hawkes processes to examine contagion in China's secondary stock market and to analyze stock market sector rotation phenomena.

\subsection{Literature Review}

This section introduces asset price predictability, temporal and spatial dependency research, and related research findings of Hawkes models in finance.

\subsubsection{Asset Price Predictability}

Whether stock prices have the nature of temporal correlation is a controversial issue. The introduction of Dow Theory in the late 19th century initiated a wave of trend analysis as technical analysis for stock investment. Corresponding to trend analysis is the random walk theory of stock prices, with \cite{bachelier1900theorie} studying the randomness of stock price changes from the perspective of Brownian motion. In 1965, economist \cite{fama1965behavior} found empirical evidence supporting the hypothesis of independent stock price increments, followed by \cite{Roberts1967} proposing three forms of efficient markets. Based on this, \citet{fama1970efficient} summarized the Efficient Market Hypothesis, stating that stock prices in an efficient market reflect all available information about assets, becoming a fundamental assumption for portfolio theory, CAPM model, option pricing, and other models.

Nevertheless, the random walk model cannot explain extraordinary stock price volatility, such as the 1987 U.S. stock market crash; much evidence suggests long-term correlation in stock prices (see \cite{mandelbrot1971can}; \cite{greene1977long}). Furthermore, \cite{jegadeesh1993returns} discovered momentum effects in stock prices, and the variance ratio test proposed by \cite{lo1989size} rejected the random walk hypothesis for stock prices. New theories continue to emerge, with \cite{peters1994fractal} proposing the Fractal Market Hypothesis based on the research of Hurst, Mandelbrot, and others, introducing chaos theory to financial markets and allowing for long-term memory. Meanwhile, \cite{delong1990noise}, \cite{lakonishok1992impact}, and others proposed theories such as herd behavior and noise traders from a behavioral finance perspective to refute the rational person assumption in the Efficient Market Hypothesis, explaining large asset price fluctuations and trend effects. Some studies indicate that China's stock market has not yet achieved weak-form efficiency (see \cite{jia2003empirical}; \cite{chen2003weak}; \cite{xiao2004empirical}). Based on the perspective that asset prices exhibit trends, this paper will study financial contagion from the correlation in the time series of the same type of events and the influence between different types of events.

\subsubsection{Characterization of Time Series, Spatial Structure, and Their Dependencies}

For time series data, many methods have been developed to handle different types of data including stationary, non-stationary, seasonal trends, and conditional heteroscedasticity, such as ARMA, ARIMA, Holt's method, GARCH, and Hidden Markov models. The Vector Autoregression (VAR) framework proposed by \cite{sims1980macroeconomics} extends autoregressive models to capture interactions between multiple time series. On the other hand, graph models and network models provide methods for capturing spatial structures and are widely applied in transportation communications, social networks, epidemiology, and other research (\cite{newman2018networks}).

After 2010, the rise of deep learning brought a data-driven perspective to capture relationships between data. Recurrent neural networks, represented by Long Short-term Memory (LSTM), are widely used in time series modeling. In terms of spatial structure, Graph Neural Networks (GNN) are powerful tools that integrate graph models with neural networks, with numerous applications in characterizing node relationships and network information transmission. Subsequently developed spatiotemporal graph models, which combine language models for processing sequential data with graph models for processing spatial structures, are considered capable of capturing both time series and spatial structure information and can be used for multivariate time series analysis (\cite{wu2020connecting}; \cite{cao2020spectral}; \cite{cheng2021modeling} used graph attention networks to model stock momentum spillover effects. However, deep learning methods face issues such as requiring large sample sizes for fitting and having weak model interpretability. In comparison, Hawkes processes model the intensity function of event occurrence, possess excellent statistical properties, and offer stronger interpretability. Multivariate Hawkes processes can also be used to capture relationships both within and between time series.

\subsubsection{Applications of Hawkes Processes in Finance}

The Hawkes process, introduced by \cite{hawkes1971spectra}, is used to characterize a class of point processes with self-exciting properties and is widely applied in high-frequency financial data analysis (\cite{bacry2015hawkes}). For example, it explains volatility clustering, with \cite{Filimonov_2012} applying it to analyze the endogenous nature of stock price volatility, forming an early warning framework for extreme events. \cite{dassios2011dynamic} extended the Hawkes and Cox models, proposing a dynamic contagion model that can capture both self-excitement and external excitement. \cite{ait2015modeling} used mutually exciting jump processes to model financial contagion. Building on the basic Hawkes model, \cite{zhu2013nonlinear} discussed the statistical properties and estimation methods of nonlinear Hawkes models. \cite{wang2016isotonic} proposed Isotonic Hawkes Processes for modeling nonlinear effects. \cite{mei2017neural} combined neural networks with Hawkes processes, using LSTM model hidden variables to drive the conditional intensity function of Hawkes processes, proposing Neural Hawkes Processes. The combination of statistical models and deep learning methods greatly enhanced the model's expressive ability, enabling it to characterize various features such as nonlinearity and self-inhibition.

Various methods have also been developed for model parameter estimation. For example, \cite{embrechts2011multivariate} adopted maximum likelihood estimation, \cite{bacry2013modelling} proposed a nonparametric estimation method based on spectral decomposition, \cite{fonseca2014hawkes} adopted generalized method of moments for parameter estimation, and \cite{xiao2017modeling} used neural networks to fit the intensity function of point processes. Given the excellent statistical properties of Hawkes models, this paper applies them to modeling contagion in China's stock market.

\subsection{Research Approach and Paper Structure}

This paper applies Hawkes processes to model financial contagion in China's stock market, assuming that asset return is driven by its own internal rules and may be influenced by price movements in other sectors. First, we use self-exciting Hawkes processes to model stock indices to test whether A-share market indices exhibit trends. Second, we extend Hawkes processes to the mutual influence between multiple sector indices to detect sector index contagion.

Part One is the introduction, presenting the research background of applying Hawkes processes to financial contagion and reviewing related domestic and international research status. Part Two covers theoretical foundations, introducing basic concepts of Hawkes processes and statistical inference methods. Part Three presents empirical analysis, describing two experiments on market index predictability and sector index contagion and their respective results. Part Four contains experimental conclusions and discussion, summarizing the conclusions drawn from the experiments.

\section{Theory}

\subsection{Hawkes Process}

\subsubsection{Basics of Hawkes Process}

The Hawkes process is a self-exciting counting process used to model the subsequent effects of extreme events. Its core concept is that the occurrence of extreme events increases the likelihood of similar events occurring afterward, similar to how aftershocks frequently follow major earthquakes. Mathematically, this influence is reflected in changes to the conditional intensity function of the counting process.

Let $\{N(t): t\geq0\}$ be a counting process. Given the history $\mathcal{H}(t)$ up to time $t$, it satisfies:
\[P(N(t+h) - N(t) = m|\mathcal{H}(t)) = \begin{cases}
1 - \lambda^*(t)h + o(h), & m = 0 \\
\lambda^*(t)h + o(h), & m = 1 \\
o(h), & m \geq 2
\end{cases}\]
where $\lambda ^*(t)$ is the conditional intensity function given the historical sequence. For convenience, the asterisk notation denotes conditioning on history, i.e., $\lambda^*(t) = \lambda(t|\mathcal{H}(u),u<t)$. The same notation applies to conditional density function $f^*(t)$ and conditional distribution function $F^*(t)$. The conditional intensity function represents the expected rate of event arrival:
\[\lambda^*(t) = \lim_{h \rightarrow 0}\frac{E(N(t + h) - N(t)|\mathcal{H}(t))}{h}.\]
\begin{definition}[Hawkes Process]
A Hawkes process is a self-exciting point process $N(t)$ with conditional intensity function $\lambda^*(t)$ defined as:
\begin{equation}
\lambda^*(t) = \mu + \int_{0}^{t}\phi(t - u)dN(u).
\end{equation}
where $\mu > 0$ is the baseline intensity, and $\phi : (0,\infty) \rightarrow [0,\infty)$ is the excitation function. The Itô integral with respect to counting process $N(u)$ is defined as:
\[\int_{0}^{t}\phi(u)dN(u) = \sum_{0 < u \leq t}\phi(u)\Delta N(u), \quad \Delta N(u) = N(u) - N(u^-).\]
\end{definition}

The excitation function $\phi(\cdot)$ is typically chosen to be exponentially decaying, such as $\phi(t) = \alpha \cdot \omega e^{-\omega(t-t_i)}$, yielding:
\begin{equation}
\lambda^*(t) = \mu + \sum_{t_i < t}\alpha \cdot \omega e^{-\omega(t - t_i)}
\end{equation}
where $\alpha > 0$ represents the jump size in intensity due to each event, and $\omega > 0$ determines the decay rate of the excitation. The stability condition for exponential kernels is $\alpha < 1$.

\paragraph{Multivariate Hawkes Process}

In multivariate Hawkes processes, events are categorized into different types, each with its own conditional intensity function that may be influenced by both its own occurrences and events of other types. For an $m$-type Hawkes process, let $\{(t_1, d_1), (t_2,d_2), \ldots, (t_{N(T)}, d_{N(T)})\}$ be the observed event sequence up to time T, where $t_i$ denotes the time of the $i$-th event and $d_i \in \{1, 2, \ldots, m\}$ represents its type.

Let $\{N_k(t): t \geq 0, k = 1, 2, \ldots, m\}$ be the point process for type $k$ events, with conditional intensity function:
\begin{equation}
\lambda_k^*(t) = \mu_k + \sum_{j = 1}^{m}\int_{0}^{t}\phi_{kj}(t - u)dN_j(u),
\end{equation}
where $\mu_k$ is the background intensity for type k events, and $\phi_{kj}(\cdot)$ is the excitation function describing the influence of type j events on type k events.
\begin{definition}[Multivariate Hawkes Process]
For an $m$-type Hawkes process, let $\{(t_1, d_1), (t_2,d_2), \\\ldots, (t_{N(T)}, d_{N(T)})\}$ be the observed event sequence, where $t_i$ denotes the time of the $i$-th event and $d_i \in \{1, 2, \ldots, m\}$ represents its type.

Let $\{N_k(t): t \geq 0, k = 1, 2, \ldots, m\}$ be the point process for type $k$ events, with conditional intensity function:
\[\lambda_k^*(t) = \mu_k + \sum_{j = 1}^{m}\int_{0}^{t}\phi_{kj}(t - u)dN_j(u)\]
where: $\mu_k > 0$ is the background intensity for type $k$ events, $\phi_{kj}(\cdot)$ is the excitation function describing the influence of type $j$ events on type $k$ events.
\end{definition}
For m-type multivariate Hawkes processes with exponential decay kernels, the vector form of the conditional intensity function is:
\begin{equation}
\bm{\lambda}^*(t) = \bm{\mu} + \sum_{(t_i,d_i):t_i < t}\bm{\alpha}_{d_i}\bm{\omega}e^{-\bm{\omega}(t - t_i)},
\end{equation}
where $\boldsymbol{\lambda}^*(t), \boldsymbol{\mu}, \boldsymbol{\alpha}_{d_i}, \boldsymbol{\omega}$ are $m \times 1 $ vectors.
The stationarity condition for a multivariate Hawkes process requires that the spectral radius of the impact matrix, denoted as $\rho(A)$, satisfies: $\rho(A) \overset{\text{def}}{=} \max_i |\lambda_i| < 1$, 
where $\lambda_i$ are the eigenvalues of the impact matrix $A$.

\begin{figure}[H]
    \centering
    \includegraphics[width=0.75\linewidth]{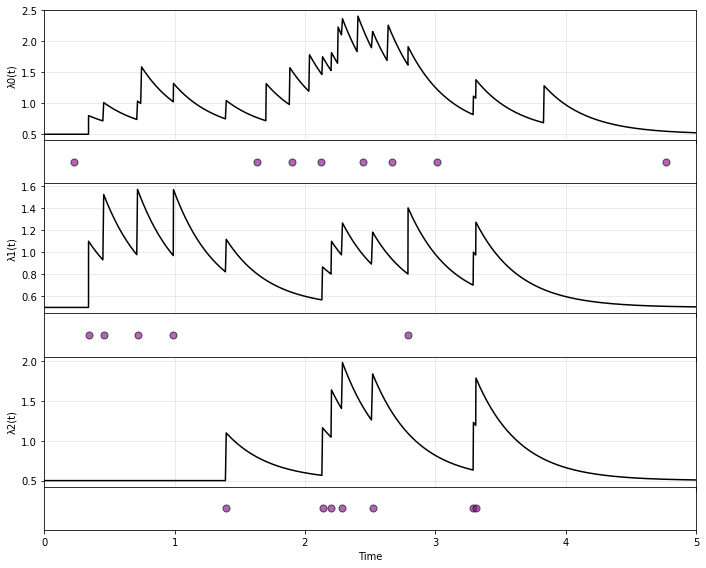}
    \caption{This figure illustrates intensity functions corresponding to a three-dimensional Hawkes process. Here, the impact matrix is configured as an upper triangle, where type 0 events are influenced by both type 1 and type 2 events. Conversely, type 2 events only receive signals from within their own type.}
\end{figure}

\paragraph{Nonlinear Hawkes Process}

To capture inhibitory effects and nonlinear influences, the Hawkes process can be extended using link functions. By employing an appropriate link function, we avoid setting constraints on $\mu$ and $\alpha$ to be positive. The conditional intensity function takes the form:
\begin{equation}
\lambda^*(t) = g(\overline{\lambda^*}(t))
\end{equation}
where $g(\cdot): \mathbb{R} \rightarrow \mathbb{R}^+$ is the link function, and $\overline{\lambda^*}(\cdot): \mathbb{R}^+ \rightarrow \mathbb{R}$ could take negative values. Common link functions include the softplus function $g(x) = \log(1+e^x)$ and the ReLU function $g(x) = \max(x,0)$.

\subsubsection{Statistical Inference for Hawkes Processes}

\paragraph{Parameter Estimation}\label{alg:pe}
This paper employs the estimation method of optimizing the likelihood function using stochastic gradient, focusing on Maximum Likelihood Estimation and stochastic gradient optimization methods.
Given event times $\{t_1, t_2, \ldots, t_{N(T)}\}$ on $[0,T]$, the likelihood function $L(\theta;T)$ for a univariate Hawkes process is:
\begin{equation}
L(\theta;T) = \left[\prod_{i = 1}^{N(T)}f^*(t_i)\right](1 - F^*(T)).
\end{equation}
Using the relationship between conditional intensity and density functions:
\begin{equation}\label{eq:int-dens}
\lambda^*(t) = \frac{f^*(t)}{1 - F^*(t)}.
\end{equation}
The likelihood function can be expressed as (shown in appendix):
\begin{equation}\label{eq:uni-ll}
L(\theta;T) = \left[\prod_{i = 1}^{N(T)}\lambda^*(t_i)\right]e^{-\int_{0}^{T}\lambda^*(s)ds}.
\end{equation}
Denote $\Lambda_j(T) = \int_0^T\lambda_j^*(s)ds$, 
for m-type multivariate Hawkes processes, the likelihood function becomes:
\begin{equation}
L(\theta;T) = \left[\prod_{i = 1}^{N(T)}\lambda_{d_i}^*(t_i)\right]e^{-\sum_{j = 1}^{m}\Lambda_j(T)}.
\end{equation}

The log-likelihood is:
\begin{equation}
\log L(\theta;T) = \sum_{i}\log\lambda_{d_i}^*(t_i) - \sum_{j = 1}^{m}\int_{0}^{t}\lambda_j^*(s)ds.
\end{equation}
The main challenge lies in computing the second term $\int_0^t \lambda_j^*(s)ds$ in the above equation. A Monte Carlo method can be employed to obtain an unbiased estimate of $\int_0^T \lambda_j^*(s)ds$, which subsequently leads to an unbiased estimate of the gradient. This is possible because:
\begin{equation}
\int_0^T \lambda_j^*(s)ds = T\int_0^T \frac{\lambda_j^*(s)}{T}ds = T\mathbb{E}_s(\lambda_j^*(s))
\end{equation}

where $\mathbb{E}_s(\cdot)$ denotes the expectation with respect to the random variable $s$ following a uniform distribution $U(0,T)$. This allows us to randomly sample $s_1,\ldots,s_N$ from $U(0,T)$, compute the mean of $\lambda_j^*(s_k)$, and update the parameters using the back-propagation (BP) algorithm by \cite{mei2017neural}. In summary, our estimation approach employs the negative log-likelihood as the objective function and utilizes stochastic gradient descent for parameter updates, as demonstrated in \ref{alg:sgd_monte_carlo}.
\begin{algorithm}[H]
\caption{Stochastic Gradient Descent Based on Monte Carlo Method}
\label{alg:sgd_monte_carlo}
\begin{algorithmic}[1]
\REQUIRE Hawkes model parameters $A$, $\mu$, $\omega$; event sequence $H(t)$; batch size $N$; observation time $T$
\FOR{each event type $m$}
    \STATE Generate $N$ samples $T_m = [t_1, t_2, \dots, t_N]$, where $t_i \sim \mathcal{U}(0, T)$.
    \STATE Compute conditional intensities $[\lambda_m^*(t_1), \lambda_m^*(t_2), \dots, \lambda_m^*(t_N)]$ using $H(t)$ and current model parameters.
    \STATE Estimate $\int_0^T \lambda_m^*(s) \, ds$ using Monte Carlo integration:
    $
    I_m = \frac{\sum_{i=1}^N \lambda_m^*(t_i)}{N} \times T
    $.
    \STATE Compute the negative log-likelihood (NLL) estimate:
    $
    \text{NLL} = -\sum_{t_i \in H(t)} \log \lambda_{d_i}^*(t_i) + \sum_{j=1}^m I_m
    $.
    \STATE Perform backpropagation: $\text{NLL.backward()}$.
    \STATE Update model parameters using gradient descent.
\ENDFOR
\ENSURE Updated model parameters $A$, $\mu$, $\omega$
\end{algorithmic}
\end{algorithm}

\section{Empirical Analysis}

\subsection{Self-Exciting Properties of Market Indices}

Experiment 1 uses daily return series of individual market indices to fit multivariate Hawkes processes, examining the self-exciting properties of daily returns for three indices: Shanghai Composite Index, Shenzhen Component Index, and ChiNext Index(Nasdaq-style Growth Enterprise Index).

\subsubsection{Dataset}
The experiment uses time series data of stock indices, including daily returns of Shanghai Composite Index, Shenzhen Component Index, and ChiNext Index. The time period spans from February 2013 to March 2023, with a sample size of 2,452. We display the descriptive statistics and cumulative return rates of those indices.

\begin{table}[h]
\caption{Descriptive Statistics of Daily Index Returns}
\begin{tabular}{lrrrrrrrrr}
\hline
\textbf{Index Name} & \textbf{Count} & \textbf{Mean} & \textbf{Std} & \textbf{Min} & \textbf{Median} & \textbf{Max} & \textbf{Skew} & \textbf{Kurtosis}\\
\hline
Shanghai Composite & 2452 & 0.021 & 1.32 & -8.491 & 0.051 & 5.764 & -0.927 & 6.71 \\
Shenzhen Component & 2452 & 0.02 & 1.601 & -8.446 & 0.036 & 6.454 & -0.683 & 3.727 \\
ChiNext & 2452 & 0.06 & 1.944 & -8.91 & 0.045 & 7.159 & -0.378 & 2.034 \\
\hline
\end{tabular}
\end{table}

\begin{figure}[h]
    \centering
    \includegraphics[width=0.9\linewidth]{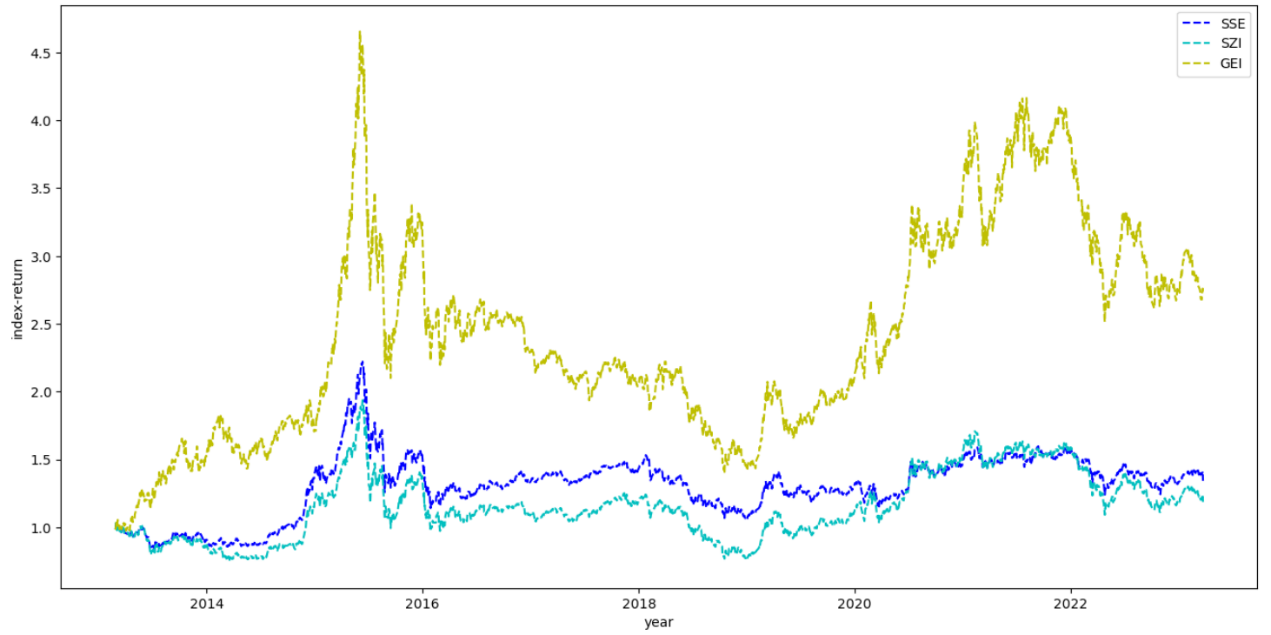}
    \vspace{12pt}  
    \caption{Cumulative return rates of major Chinese market indices (Shanghai Stock Exchange Composite Index(SSE), Shenzhen Component Index(SZI), and Growth Enterprise Index(GEI)) from February 26, 2013 to March 2023. The base value is normalized to 1.0 on February 26, 2013.}
    \label{fig:market-indices-returns}
\end{figure}
During the study period, Shenzhen Component Index and ChiNext Index showed higher volatility. ChiNext Index returns exhibited characteristics of high mean and high volatility, while Shanghai Composite Index displayed left-skewed, leptokurtic features. The statistical properties indicate different market styles, closely related to their constituent stocks. The Shanghai Composite Index(SSE) mainly comprises traditional industries such as finance, real estate, and consumption, while Shenzhen Component(SZI) and ChiNext(GEI) indices include many healthcare and information technology companies.
\subsubsection{Experimental Design and Results}

Following \cite{embrechts2011multivariate}, daily returns were divided into three segments based on the 10th and 90th percentiles of the Shanghai Composite Index return history: sharp decline, normal range, and sharp rise. When returns fall into either the sharp decline or sharp rise range, it is recorded as a special event occurrence. The return series was transformed into sequences of two types of event occurrences, fitting a bivariate Hawkes model. 85\% of the time series was used as the training set (2,084 samples) and 15\% as the validation set (368 samples). Sharp rise events were labeled as type 0, sharp decline events as type 1. An exponential kernel Hawkes model was fitted using Python's tick module for parameter estimation, yielding:

Impact matrix and baseline intensity:
\begin{equation}
A = [\alpha_{ij}] = \begin{bmatrix}
0.382 & 0.387 \\
0.218 & 0.343
\end{bmatrix}, \ \mu = [0.024 \quad 0.044]
\end{equation}

The decay rate $\omega$ was set as a hyperparameter, assuming consistent decay rates for both event types. Grid search determined $\omega = 0.1$.
In the impact matrix, $\alpha_{ij}$ represents the influence of the occurrence of the jth type of event on the intensity function of the ith type of event. It is noted that regardless of the occurrence of significant upward events (first column of the impact matrix) or significant downward events (second column of the impact matrix), the impact on the conditional intensity function of significant upward events is higher than that of significant downward events. This indicates that the likelihood of the stock index maintaining growth after a significant increase is higher than after a significant decrease, while there is a greater possibility of a rebound after a significant decrease; however, the possibility of further decline is also considerable. Among the four values in the impact matrix, only the transmission impact of significant upward events on significant downward events is relatively small, which to some extent reflects the rationality of chasing gains and cutting losses. The relatively large values of $\alpha_{00}$ and $\alpha_{11}$ demonstrate the continuity of upward or downward trends. Additionally, the relatively large value of $\alpha_{01}$ indicates that the Shanghai Composite Index is prone to experiencing "oversold rebounds." The conditional intensity functions of the two types of events are depicted below.
\begin{figure}[H]
    \centering
    \includegraphics[width=0.85\linewidth]{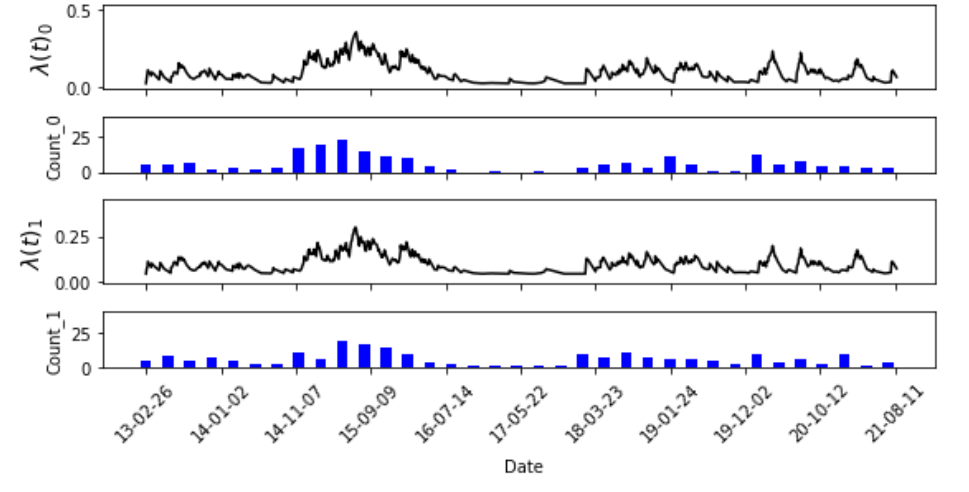}\caption{Conditional density functions and barplots of defined extreme events in the SSE Index (2013-2021), where Type 0 and Type 1 represent large upward and downward movements, respectively. The market was very active in 2014-2015, followed by a period of relative stability from 2016 to 2018.}
    \label{fig:enter-label}
\end{figure}

Using the same method for Shenzhen Component Index and ChiNext Index daily returns, the quantiles and estimates are:

\begin{table}[H]
\caption{Parameter Comparison of Three Major Indices}
\begin{tabular}{lllll}
\hline
Index Name & 0.1, 0.9 quantile (\%) & $\omega$ & Baseline $\mu$ & Impact Matrix $A$ \\
\hline
Shanghai Composite & [-1.285 1.481] & 0.1 & $[0.024 ; 0.044]$ & $\begin{bmatrix} 0.382 & 0.387 \\ 0.218 & 0.343 \end{bmatrix}$ \\
\hline
Shenzhen Component & [-1.701 1.891] & 0.1 & $[0.038 ; 0.047]$ & $\begin{bmatrix} 0.175 & 0.451 \\ 0.155 & 0.381 \end{bmatrix}$ \\
\hline
ChiNext & [-2.105 2.502] & 0.1 & $[0.034 ; 0.05]$ & $\begin{bmatrix} 0.28 & 0.385 \\ 0.154 & 0.348 \end{bmatrix}$ \\
\hline
\end{tabular}
\end{table}

\subsection{Hawkes Process Variants: Inhibition and Nonlinear Effects}

A variant of Hawkes processes allows for self-inhibition or mutual inhibition characteristics, for example, by allowing $\alpha_{ij} < 0$ in the exponential kernel form while ensuring the conditional intensity function $\lambda^*(t)$ remains non-negative. A more generalized form allows event impacts to propagate nonlinearly through link functions. Experiment 2 uses extended Hawkes processes to explore market sector rotation effects.

\subsubsection{Dataset}

Experiment 2 data includes three sector indices: CSI Consumer Index (000932), CSI Healthcare Index (000933), and CSI Finance \& Real Estate Index (000934) daily return series, spanning from March 2013 to March 2023, with 2,452 samples.

\begin{table}[H]
\caption{Descriptive Statistics of Sector Index Daily Returns}
\begin{tabular}{lrrrrrrrr}
\hline
\textbf{Index Name} & \textbf{Count} & \textbf{Mean} & \textbf{Std} & \textbf{Min} & \textbf{Median} & \textbf{Max} & \textbf{Skew} & \textbf{Kurtosis}\\
\hline
CSI Consumer & 2452 & 0.067 & 1.648 & -8.325 & 0.061 & 7.214 & -0.361 & 2.753 \\
CSI Healthcare & 2452 & 0.035 & 1.659 & -8.484 & 0.067 & 7.909 & -0.395 & 2.67 \\
CSI Fin\&RE & 2452 & 0.023 & 1.557 & -9.363 & -0.034 & 8.918 & -0.086 & 5.173 \\
\hline
\end{tabular}
\end{table}

\begin{figure}[H]
    \centering
    \includegraphics[width=0.85\linewidth]{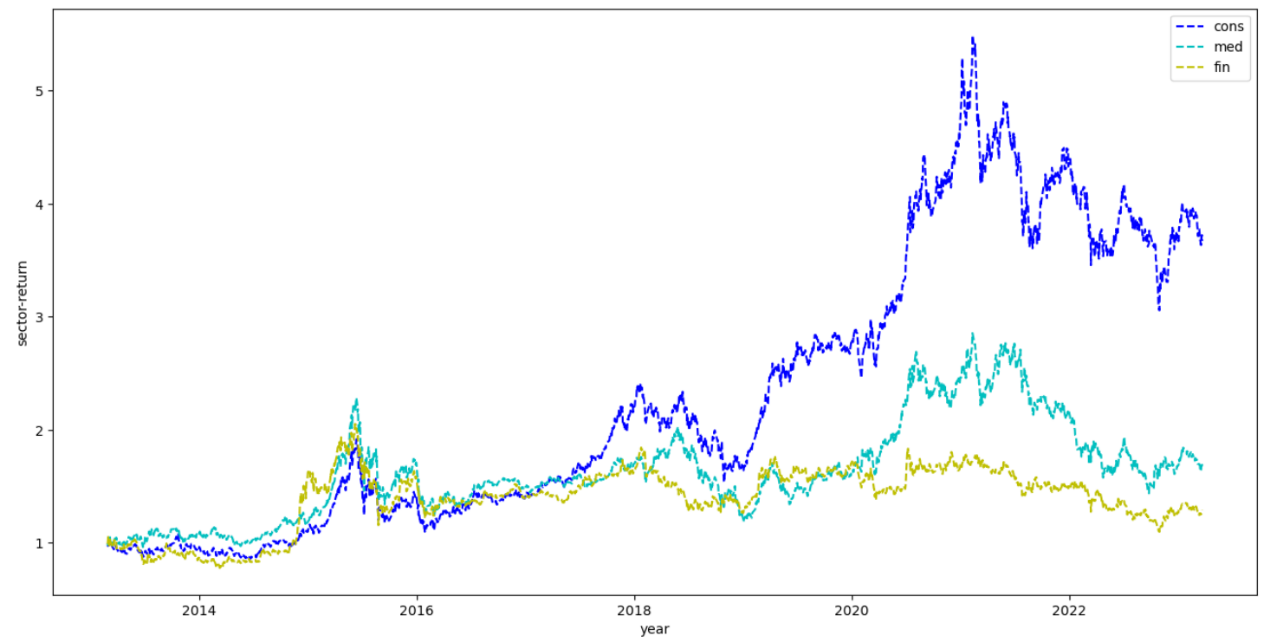}
    \caption{Normalized returns across three sectors (consumption, medical industry, and financial) in Chinese stock markets, 2013-2023. The baseline return is normalized to unity on February 26, 2013. The figure illustrates the relative performance and temporal evolution of these sectors over a decade-long period.}
    \label{fig:enter-label}
\end{figure}
    
\subsubsection{Experimental Design and Results}

Given the time-sensitive nature of sector rotation, we initially segmented the sample sequence into 15 periods, with each period comprising 150 trading days. Following previous conventions, extreme upward and downward movements are considered as two distinct event types. With three sectors under consideration, this results in six different event categories, enabling us to model the contagion effects using a six-dimensional Hawkes process.
We assume the conditional intensity function takes the form:
\begin{equation}\label{eq:mult}
\lambda^*(t) = g(\overline{\lambda^*}(t)). \ \overline{\lambda^*}(t) = \mu + \alpha\omega e^{-\omega t}, \mu \in \mathbb{R},\alpha \in \mathbb{R}, \omega > 0. \
g(x) = \max(x,0.01).
\end{equation}

In this setting, we remove the non-negativity constraints on both the background intensity $\mu$ and the influence matrix parameter $\alpha$ to model inhibitory effects. The link function $g(\cdot)$ ensures the non-negativity of conditional intensity.

Parameter estimation is conducted via algorithms presented in section \ref{alg:pe}. Representative periods 3(2014.12 to 2015.08) and periods 5(2016.03 to 2016.11) are selected for analysis based on market activity levels:

\begin{itemize}
\item Period 3 (2014-12 to 2015-08): a structural bull market with higher frequencies of significant upward movements compared to downward movements.

\item Period 5 (2016-03 to 2016-11): pessimistic sentiment, characterized by reduced trading activity and notably lower frequencies of both substantial gains and losses compared to previous periods.
\end{itemize}



The estimated model parameters are given in table \ref{tab:mult-params}.
\begin{table}[ht]
\centering
\caption{Estimated parameters for the Hawkes process model \eqref{eq:mult}: activation rate ($A_{ij}$), background intensity ($\mu$), and decay rate ($\omega$) for Period 3 and Period 5. Category 0-5 represents event type: cons-up, med-up, fin-up, cons-down, med-down, fin-down respectively. The element $A_{ij}$ represents the excitation effect of type j 
event on type i.}

\label{tab:mult-params}
\small
\setlength{\tabcolsep}{2pt}

\begin{subtable}{0.48\textwidth}
\centering
\caption{Influence matrix of Period 3(2014.12-2015.08) }
\begin{tabular}{@{}ccccccc@{}}
\hline
$A_{ij}$ & 0 & 1 & 2 & 3 & 4 & 5 \\
\hline
0 & 1.161 & 0.240 & 0.253 & -0.335 & -0.469 & -0.154 \\
1 & 0.157 & 1.143 & -0.238 & -0.230 & -0.173 & -0.208 \\
2 & -0.105 & -0.298 & 1.313 & -0.371 & -0.474 & -0.425 \\
3 & 0.109 & 0.093 & 0.002 & 0.671 & 0.420 & -0.067 \\
4 & 0.037 & -0.191 & -0.005 & 0.460 & 0.831 & 0.078 \\
5 & -0.366 & -0.100 & -0.136 & 0.071 & 0.033 & 1.092 \\
\hline
\end{tabular}
\end{subtable}
\hfill
\begin{subtable}{0.48\textwidth}
\centering
\caption{Influence matrix of Period 5(2016.03-2016.11)}
\begin{tabular}{@{}ccccccc@{}}
\hline
$A_{ij}$ & 0 & 1 & 2 & 3 & 4 & 5 \\
\hline
0 & -0.317 & -0.536 & -0.323 & -0.340 & -0.424 & -0.452 \\
1 & -0.375 & -0.437 & -0.315 & -0.478 & -0.379 & -0.600 \\
2 & -0.564 & -0.314 & -0.136 & -0.408 & -0.574 & -0.577 \\
3 & -0.576 & -0.452 & -0.427 & -0.265 & -0.470 & -0.517 \\
4 & -0.405 & -0.505 & -0.396 & -0.406 & -0.346 & -0.401 \\
5 & -0.203 & -0.535 & -0.418 & -0.334 & -0.403 & -0.338 \\
\hline
\end{tabular}
\end{subtable}

\vspace{1cm}

\begin{subtable}{0.95\linewidth}
\centering
\caption{Background Intensity $\mu$ and Decay Parameter $\omega$}
\label{tab:background_parameters}
\begin{tabular}{cll}
\hline
Period & \multicolumn{1}{c}{$\mu$} & $\omega$ \\
\hline
3 & [-0.226, -0.200, -0.341, -0.079, -0.024, -0.095] & 0.702 \\
5 & [0.089, 0.093, 0.058, 0.091, 0.091, 0.061] & 0.004 \\
\hline
\end{tabular}
\end{subtable}
\end{table}

The $(6 \times 6)$ influence matrix, partitioned into four $(3 \times 3)$ blocks
$
\begin{bmatrix} 
A_{3\times 3} & B_{3\times 3} \\
C_{3\times 3} & D_{3\times 3}
\end{bmatrix}
$. Block $A$ indicates upward trend; Block $B$ represents oversold rebound; Block $C$ manifests pullback and Block $D$ displays the downward trend. 

During active trading periods, such as period 3, the generalized Hawkes model identified evidence of both upward and downward trend continuity, which is underscored by the presence of diagonal elements in the estimated influence matrix.

Comparison of off-diagonal blocks exposes cross-trend influences. Greater intensity within these blocks suggests a swifter rotation of market focus. In contrast, during market dormancy, exemplified by period 5, the disparity between diagonal and off-diagonal values diminishes, suggesting potential sector rotation phenomenon. A reduced decay parameter during a stagnant market phase signifies market inertia.


\section{Conclusions}
In the first experiment, this study examined the self-exciting characteristics of three major market indices—Shanghai Composite Index, Shenzhen Component Index, and ChiNext Index—through multivariate Hawkes processes. The evidence supported the existence of both upward and downward trends in the Shanghai Composite Index, as well as oversold rebounds.

In the second experiment, we modeled the contagion effects among Consumption, Healthcare, and Financial sector indices using Hawkes processes that incorporated inhibition and nonlinear effects. The findings revealed that Hawkes processes can effectively investigate market styles across different periods. During the high-trading-volume period from January 2015 to March 2016, trend continuation was evident in both structural bull and bear trend. Conversely, during the low-trading-volume period from March 2016 to August 2016, the market exhibited sector rotation patterns.

\bibliographystyle{aer}
\bibliography{biblio}

@article{chen2003weak,
    title={Is Chinese Stock Market Weakly Efficient?},
    author={Chen, Dengta and Hong, Yongmiao},
    journal={China Economic Quarterly},
    volume={3},
    number={1},
    pages={97--124},
    year={2003}
}

@article{jia2003empirical,
    title={Empirical Analysis of Chinese Stock Market Efficiency},
    author={Jia, Quan and Chen, Zhangwu},
    journal={Journal of Financial Research},
    volume={7},
    pages={86--92},
    year={2003}
}

@article{xiao2004empirical,
    title={Empirical Study on the Effectiveness of Value Reversal Investment Strategy in Chinese Stock Market},
    author={Xiao, Jun and Xu, Xinzhong},
    journal={Economic Research Journal},
    volume={3},
    pages={55--64},
    year={2004}
}

@article{ait2015modeling,
    title={Modeling financial contagion using mutually exciting jump processes},
    author={A\"it-Sahalia, Y and Cacho-Diaz, J and Laeven, R J A},
    journal={Journal of Financial Economics},
    volume={117},
    number={3},
    pages={585--606},
    year={2015}
}

@article{bachelier1900theorie,
    title={Theorie de la speculation},
    author={Bachelier, L},
    journal={Ann. Sci. Ecole Norm. Sup.},
    volume={17},
    pages={21--86},
    year={1900}
}

@article{bacry2013modelling,
    title={Modelling microstructure noise with mutually exciting point processes},
    author={Bacry, E and Delattre, S and Hoffmann, M and others},
    journal={Quantitative finance},
    volume={13},
    number={1},
    pages={65--77},
    year={2013}
}

@article{bacry2015hawkes,
    title={Hawkes processes in finance},
    author={Bacry, E and Mastromatteo, I and Muzy, J F},
    journal={Market Microstructure and Liquidity},
    volume={1},
    number={01},
    pages={1550005},
    year={2015}
}

@article{cao2020spectral,
    title={Spectral temporal graph neural network for multivariate time-series forecasting},
    author={Cao, D and Wang, Y and Duan, J and others},
    journal={Advances in neural information processing systems},
    volume={33},
    pages={17766--17778},
    year={2020}
}

@inproceedings{cheng2021modeling,
    title={Modeling the momentum spillover effect for stock prediction via attribute-driven graph attention networks},
    author={Cheng, R and Li, Q},
    booktitle={Proceedings of the AAAI Conference on artificial intelligence},
    volume={35},
    number={1},
    pages={55--62},
    year={2021}
}

@article{dassios2011dynamic,
    title={A dynamic contagion process},
    author={Dassios, A and Zhao, H},
    journal={Advances in applied probability},
    volume={43},
    number={3},
    pages={814--846},
    year={2011}
}

@article{delong1990noise,
    title={Noise trader risk in financial markets},
    author={De Long, J B and Shleifer, A and Summers, L H and others},
    journal={Journal of political Economy},
    volume={98},
    number={4},
    pages={703--738},
    year={1990}
}

@article{embrechts2011multivariate,
    title={Multivariate Hawkes processes: an application to financial data},
    author={Embrechts, P and Liniger, T and Lin, L},
    journal={Journal of Applied Probability},
    volume={48},
    number={A},
    pages={367--378},
    year={2011}
}

@article{fama1970efficient,
    title={Efficient capital markets: A review of theory and empirical work},
    author={Fama, E F},
    journal={The journal of Finance},
    volume={25},
    number={2},
    pages={383--417},
    year={1970}
}

@article{fama1965behavior,
    title={The behavior of stock-market prices},
    author={Fama, E F},
    journal={The journal of Business},
    volume={38},
    number={1},
    pages={34--105},
    year={1965}
}

@article{greene1977long,
    title={Long-term dependence in common stock returns},
    author={Greene, M T and Fielitz, B D},
    journal={Journal of Financial Economics},
    volume={4},
    number={3},
    pages={339--349},
    year={1977}
}

@article{hawkes1971spectra,
    title={Spectra of some self-exciting and mutually exciting point processes},
    author={Hawkes, A G},
    journal={Biometrika},
    volume={58},
    number={1},
    pages={83--90},
    year={1971}
}

@article{jegadeesh1993returns,
    title={Returns to buying winners and selling losers: Implications for stock market efficiency},
    author={Jegadeesh, N and Titman, S},
    journal={The Journal of finance},
    volume={48},
    number={1},
    pages={65--91},
    year={1993}
}

@article{lakonishok1992impact,
    title={The impact of institutional trading on stock prices},
    author={Lakonishok, J and Shleifer, A and Vishny, R W},
    journal={Journal of financial economics},
    volume={32},
    number={1},
    pages={23--43},
    year={1992}
}

@article{lo1989size,
    title={The size and power of the variance ratio test in finite samples: A Monte Carlo investigation},
    author={Lo, A W and MacKinlay, A C},
    journal={Journal of econometrics},
    volume={40},
    number={2},
    pages={203--238},
    year={1989}
}

@article{mandelbrot1971can,
    title={When can price be arbitraged efficiently? A limit to the validity of the random walk and martingale models},
    author={Mandelbrot, B B},
    journal={The Review of Economics and Statistics},
    pages={225--236},
    year={1971}
}

@article{mei2017neural,
    title={The neural hawkes process: A neurally self-modulating multivariate point process},
    author={Mei, H and Eisner, J M},
    journal={Advances in neural information processing systems},
    volume={30},
    year={2017}
}

@book{newman2018networks,
    title={Networks},
    author={Newman, M},
    publisher={Oxford university press},
    year={2018}
}

@book{peters1994fractal,
    title={Fractal market analysis: applying chaos theory to investment and economics},
    author={Peters, E E},
    publisher={John Wiley \& Sons},
    year={1994}
}

@article{sims1980macroeconomics,
    title={Macroeconomics and reality},
    author={Sims, C A},
    journal={Econometrica: journal of the Econometric Society},
    pages={1--48},
    year={1980}
}

@inproceedings{wang2016isotonic,
    title={Isotonic hawkes processes},
    author={Wang, Y and Xie, B and Du, N and others},
    booktitle={International conference on machine learning},
    pages={2226--2234},
    organization={PMLR},
    year={2016}
}

@inproceedings{wu2020connecting,
    title={Connecting the dots: Multivariate time series forecasting with graph neural networks},
    author={Wu, Z and Pan, S and Long, G and others},
    booktitle={Proceedings of the 26th ACM SIGKDD international conference on knowledge discovery \& data mining},
    pages={753--763},
    year={2020}
}

@inproceedings{xiao2017modeling,
    title={Modeling the intensity function of point process via recurrent neural networks},
    author={Xiao, S and Yan, J and Yang, X and others},
    booktitle={Proceedings of the AAAI conference on artificial intelligence},
    volume={31},
    number={1},
    year={2017}
}

@phdthesis{zhu2013nonlinear,
    title={Nonlinear Hawkes processes},
    author={Zhu, L},
    school={New York University},
    year={2013}
}

@article{Filimonov_2012,
   title={Quantifying reflexivity in financial markets: Toward a prediction of flash crashes},
   volume={85},
   ISSN={1550-2376},
   url={http://dx.doi.org/10.1103/PhysRevE.85.056108},
   DOI={10.1103/physreve.85.056108},
   number={5},
   journal={Physical Review E},
   publisher={American Physical Society (APS)},
   author={Filimonov, Vladimir and Sornette, Didier},
   year={2012},
   month=may }

@article{fonseca2014hawkes,
    title={Hawkes Process: Fast Calibration, Application to Trade Clustering, and Diffusive Limit},
    author={Fonseca, Jos{\'e} Da and Zaatour, Riadh},
    journal={Journal of Futures Markets},
    volume={34},
    number={6},
    pages={548--579},
    year={2014},
    doi={10.1002/fut.21644}
}

@unpublished{Roberts1967,
  author = {Roberts, Harry V.},
  year = {1967},
  title = {Statistical versus Clinical Prediction of the Stock Market},
  note = {Unpublished manuscript, Center for Research in Security Prices, University of Chicago}
}

\appendix

\section{Supplementary Derivations for Statistical Inference of Hawkes Process in Chapter 2, Section 2}
\subsection{Proof of $\lambda^{*}(t) = \frac{f^{*}(t)}{1 - F^{*}(t)}.$}
\begin{proof}[Proof of equation \eqref{eq:int-dens}]
Since $P(N(t+h)-N(t)\geq2|\mathcal{H}(t)) = o(h)$, and assuming $i-1$ events have occurred at time $t$, i.e., $N(t) = i - 1$, and letting $X_{(i)}$ denote the occurrence time of the $i$-th event, we have:
\begin{equation}
\begin{aligned}
\lambda^{*}(t) &= \lim_{h \rightarrow 0}\frac{E(N(t + h) - N(t)|\mathcal{H(}t))}{h} \\
&= \lim_{h \rightarrow 0}\frac{P(X_{i} \in (t,t + h)|X_{i} > t)}{h} \\
&= \lim_{h \rightarrow 0}\frac{F^{*}(t + h) - F^{*}(t)}{\left( 1 - F^{*}(t) \right)h} \\
&= \frac{f^{*}(t)}{1 - F^{*}(t)}
\end{aligned}
\end{equation}
\end{proof}

\subsection{Proof of $f^{*}(t_{i}) = \lambda^{*}(t_{i})\exp (-\int_{t_{i-1}}^{t_{i}}\lambda^{*}(s)ds)$}
\begin{proof}
From the result in Appendix A.1, $\lambda^{*}(t) = \frac{f^{*}(t)}{1 - F^{*}(t)} = - d\ln(1 - F^{*}(t))$. Integrating t from $t_{i-1}$ to $t_i$, we have:

\begin{equation}
\int_{t_{i-1}}^{t_{i}}\lambda^{*}(t)dt = -(ln(1-F^{*}(t_i))-ln(1-F^{*}(t_{i-1})))
\end{equation}

Under the simple point process assumption, the probability of multiple events occurring at the same instant is zero, thus $F^{*}(t_{i-1})=0$. Therefore:
\begin{equation}
\int_{t_{i-1}}^{t_{i}}\lambda^{*}(t)dt = -ln(1-F^{*}(t_i))
\end{equation}

Substituting $$1-F^{*}(t_i) = \lambda^{*}(t_i)f^{*}(t_i),$$ we obtain:
\begin{equation}\label{eq:apd1}
f^{*}(t_i) = \lambda^{*}(t_i)\exp(-\int_{t_{i-1}}^{t_i}\lambda^{*}(s)ds)
\end{equation}
\end{proof}

\subsection{Proof of $L(\theta;T) = \left[\prod_{i=1}^{n(T)}\lambda^{*}(t_i)\right]e^{-\int_0^T\lambda^{*}(s)ds}$}
\begin{proof}[Proof of equation \eqref{eq:uni-ll}]
Since
\begin{equation*}
L(\theta;T) = \left[\prod_{i=1}^{N(T)}f^{*}(t_i)\right](1-F^{*}(T)) 
\ \ \text{and}\ \  1-F^{*}(T) = \exp(-\int_{t_{N(T)}}^T\lambda^{*}(s)ds)
\end{equation*}
Substituting the result from equation \eqref{eq:apd1}:
\begin{equation}
\begin{aligned}
f^{*}(t_i) = \left[\prod_{i=1}^{N(T)}\lambda^{*}(t_i)\right]  
\exp\left(-\sum_{i=1}^{N(T)}\int_{t_{i-1}}^{t_i}\lambda^{*}(s)ds - \int_{t_{N(T)}}^T\lambda^{*}(s)ds\right)
\end{aligned}
\end{equation}
Therefore:
\begin{equation}
L(\theta;T) = \left[\prod_{i=1}^{N(T)}\lambda^{*}(t_i)\right]e^{-\int_0^T\lambda^{*}(s)ds}
\end{equation}
\end{proof}

\section{Goodness-of-Fit Testing}

\begin{theorem}[Random Time Change Theorem]\label{thm:time-change}
Given a point process realization $\{t_1, t_2, \ldots, t_k\}$ on $[0,T]$, if the conditional intensity function satisfies $\lambda^*(t) > 0$, $t \in [0, T]$ and compensator $\Lambda(T) = \int_0^T \lambda^*(s)ds < \infty$ almost everywhere, then the time points $\{\Lambda(t_1), \Lambda(t_2), \ldots, \Lambda(t_k)\}$ follow a unit-rate Poisson process. Conversely, $\{t_1, t_2, \ldots, t_k\}$ is a realization of a point process with conditional intensity function $\lambda^*(t)$ if and only if $\{\Lambda(t_1), \Lambda(t_2), \ldots, \Lambda(t_k)\}$ follows a unit-rate Poisson process.
\end{theorem}

If the $\lambda^*(t)$ is correctly specified, the sequence of compensator ${\Lambda(t_i)}$ follows a unit-rate Poisson process. For multivariate Hawkes processes, we can verify whether the transformed sequences $\{\Lambda_k(t_1), \Lambda_k(t_2), \ldots, \Lambda_k(t_k)\}$, $k = 1, 2, \ldots, m$, follow unit-rate Poisson processes. Q-Q plots can be used to check the relationship between empirical and theoretical quantiles. For testing independence in Poisson processes, we can examine the scatter plot of transformed intervals $\tau_{i+1} - \tau_i$, where $\tau_i = \Lambda(t_{i+1}) - \Lambda(t_i)$. The absence of obvious patterns suggests temporal independence.

In the fitting tests, a homogeneous Poisson process was used as the baseline model to fit the sequence data, comparing results with the Hawkes process.

(A) Homogeneous Poisson Process Fitting and Testing

Under the homogeneous Poisson process assumption, inter-event times follow an exponential distribution. Using moment estimation $\hat{\lambda} = \frac{1}{\sum_{i}\frac{\Delta t_{i}}{n}}$
where $\Delta t_i$ is the $i$th interval time, $n$ is the total number of events. For period between 2013.02 and 2022.03, 
$$\hat{\lambda}_{train} = [0.100, 0.101],$$
For period between 2022.03 and 2023.03:
$\hat{\lambda}_{test} = [0.067, 0.082].$
$\hat{\lambda}_{test} < \hat{\lambda}_{train}$ indicates that the average interval between extreme events has increased in the past year, suggesting a scheme change in market structure.

\begin{figure}[H]
    \centering
    \includegraphics[width=0.85\linewidth]{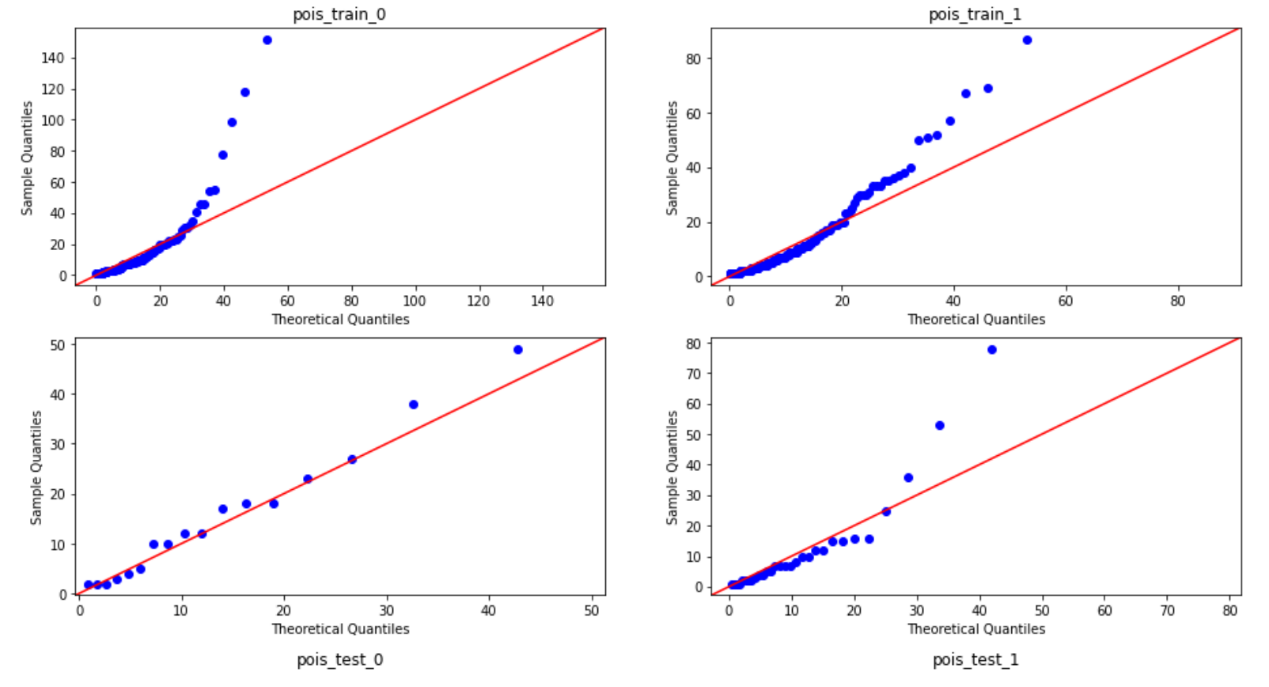}
    \caption{Q-Q Plot under homogeneous Poisson assumption.}
    \label{fig:enter-label}
\end{figure}

(B) Hawkes Process Fitting Tests

According to Theorem \ref{thm:time-change}, Q-Q plots were used to test whether time intervals $\Delta\Lambda(t_i)$ follow an exponential distribution with scale parameter 1.

\begin{figure}[H]
    \centering
    \includegraphics[width=0.85\linewidth]{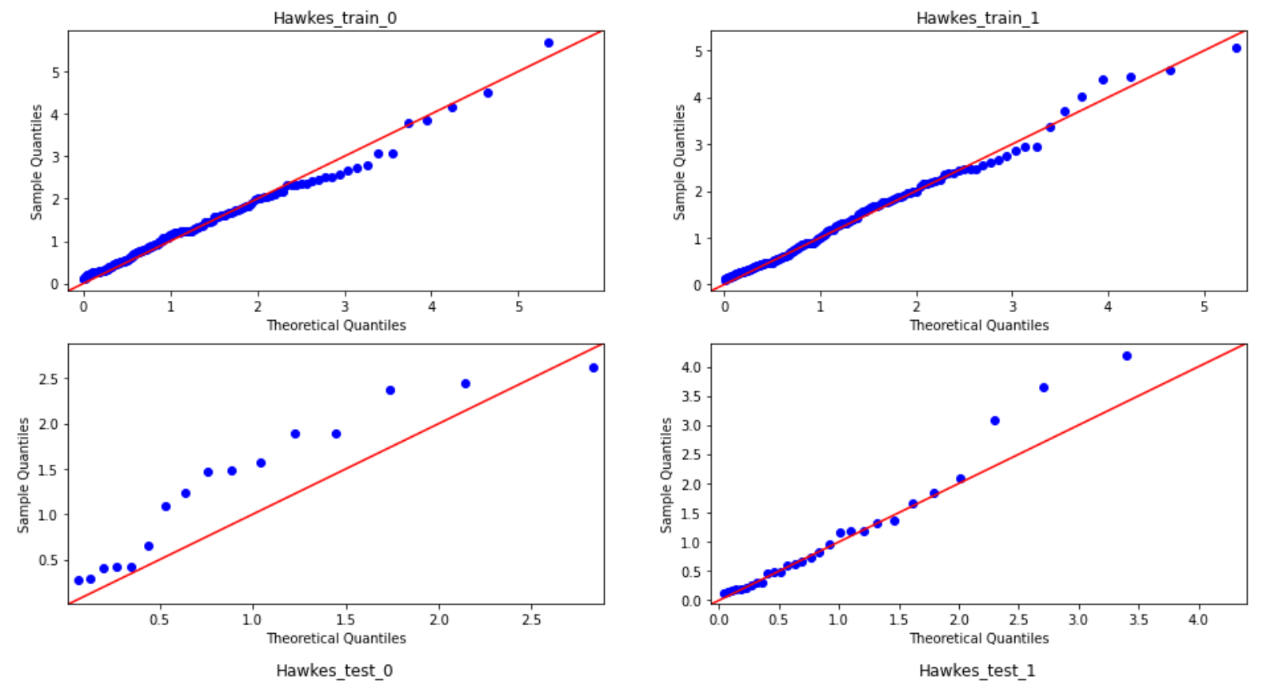}
    \caption{Q-Q Plot under Hawkes process assumption.}
\end{figure}

\begin{figure}[H]
    \centering
    \includegraphics[width=0.85\linewidth]{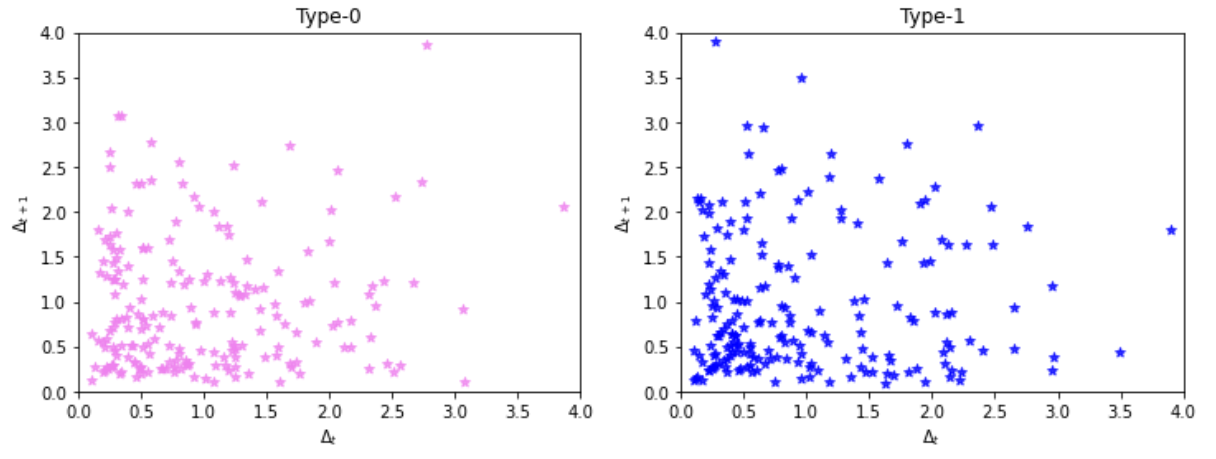}
    \caption{Scatter plot of transformed inter-event time intervals $(\Lambda_{t+1}-\Lambda_t, \Lambda_t - \Lambda_{t-1})$. The absence of discernible patterns suggests the statistical independence of consecutive time intervals, as expected under the time transformation theorem for point processes.}
\end{figure}

\end{document}